\documentclass[10pt,twocolumn,pra,aps,superscriptaddress]{revtex4-1}
\usepackage{amssymb,amsmath,amsfonts}

\usepackage[latin9]{inputenc}
\usepackage{color}
\usepackage{amstext}
\usepackage{enumitem}
\usepackage{graphicx}
\usepackage[colorlinks=true,linkcolor=blue,urlcolor=blue,citecolor=blue,pdfusetitle]{hyperref}

\usepackage[sc]{mathpazo}

\usepackage{hyperref,cleveref}

\usepackage[dvipsnames]{xcolor}
\usepackage[caption=false]{subfig}

\makeatletter

\newcommand\ket[1]{\left|#1\right\rangle}
\newcommand\bra[1]{\left\langle #1 \right|}

\usepackage{times}
\usepackage{bbm}

\makeatother

\begin{document}
\title{On the role of initial coherence in the spin phase-space entropy production rate}
\date{\today}

\author{Giorgio Zicari}\email{gzicari01@qub.ac.uk}
\affiliation{Centre for Quantum Materials and Technologies, School of Mathematics and Phyiscs, Queen's University, Belfast BT7 1NN, United Kingdom}

\author{Bar{\i}\c{s} \c{C}akmak} 
\affiliation{College of Engineering and Natural Sciences, Bah\c{c}e\c{s}ehir University, Be\c{s}ikta\c{s}, Istanbul 34353, T\"urkiye}
\affiliation{TUBITAK Research Institute for Fundamental Sciences, 41470 Gebze, T\"urkiye}

\author{\"Ozg\"ur E. M\"ustecapl{\i}o\u{g}lu}
\affiliation{Department of Physics, Ko\c{c} University, \.{I}stanbul, 
Sar{\i}yer, 34450, T\"urkiye}
\affiliation{TUBITAK Research Institute for Fundamental Sciences, 41470 Gebze, T\"urkiye}

\author{Mauro Paternostro}
\affiliation{Centre for Quantum Materials and Technologies, School of Mathematics and Phyiscs, Queen's University, Belfast BT7 1NN, United Kingdom}

\begin{abstract}
{Recent studies have pointed out the intrinsic dependence of figures of merit of thermodynamic relevance -- such as work, heat and entropy production --  on the amount of quantum coherences that is made available to a system. However, whether coherences hinder or enhance the value taken by such quantifiers of thermodynamic performance is yet to be ascertained. We show that, when considering entropy production generated in a process taking a finite-size bipartite quantum system out of equilibrium through local non-unitary channels, no general monotonicity relationship exists between the entropy production and degree of quantum coherence in the state of the system. A direct correspondence between such quantities can be retrieved when considering specific forms of open-system dynamics applied to suitably chosen initial states. Our results call for a systematic study of the role of genuine quantum features in the non-equilibrium thermodynamics of quantum processes.}
\end{abstract}

\maketitle

\section{Introduction}
\label{sec:intro}

The growing interest towards quantum technologies has fostered a variety of different theoretical considerations aiming at assessing the potential advantages of quantum resources with respect to classical ones. Some of them fall into the domain of quantum thermodynamics, which provides a framework to explore the validity of thermodynamics laws in the quantum regime, not only at a fundamental level, but also in terms of their technological applications~\cite{ThermoBook:2018, Goold_2016, DeffnerCampbellBook,Myers:2022}. 

In this respect, this theoretical effort comes up against the impossibility of modelling quantum systems as perfectly closed and isolated from their surroundings. Therefore, the theory of open quantum systems appears to be a natural candidate to model and interpret a number of quantum thermodynamics problems~\cite{Breuer-Petruccione, Alicki:2007, Rivas2012}. 
For instance, a long-standing problem is how to state and consistently interpret the second law of thermodynamics in the quantum domain. While moving within the weak-coupling and memoryless description of the dynamics, i.e. the so-called Born-Markov approximation, one can restate the second law of thermodynamics by resorting to the formalism offered by the theory of open quantum systems. The situation becomes more problematic in those regimes in which the Born-Markov approximation breaks down, leading to memory or strong coupling effects~\cite{Strasberg:2016,Perarnau-Llobet:2018,Jen-Tsung:2018,Rivas:2020}: even the definitions of work and heat need to be carefully discussed.

Beside the Clausius and Kelvin-Planck's statements, the second law of thermodynamics can be formulated in terms of entropy, or, in non-equilibrium scenarios, in terms of entropy production rate, the latter accounting for the rate with which the entropy is intrinsically produced by the physical processes taking place within the system~\cite{Mauro_review}. One can indeed show that, for classical systems, this quantity always needs to be non-negative~\cite{DeGrootMazur}. Upon a suitable identification of heat and work, this fundamental result can be recovered in the quantum regime, provided that one considers the standard Born-Markov scenario, as first rigorously shown by Spohn~\cite{Spohn:1978}. In contrast, if the system's dynamics is resolved on a timescale such that non-Markovian effects cannot be completely washed out, there might be time intervals in which the entropy production rate gets negative values~\cite{Marcantoni:2017, PhysRevA.95.012122,Popovic:2018,StrasbergPRE:2019,SenyasaEntropy:2022}. As counter-intuitive as it might seem at first glance, this occurrence can be framed in the picture of non-Markovianity as backflow of information: in the usual system-environment scenario, the system can partially recover the information that was previously lost due to its interaction with a much larger environment~\cite{Breuer:16,Rivas:14,deVega:2017}.

Entropy production plays a crucial role both in the classical and quantum domains: it is a crucial factor in the determination of the efficiency of a thermal engine and, therefore, its minimisation is desirable to get closer to the ideal Carnot bound. It is endowed with fundamental relevance, as it captures some of the features witnessing the irreversibilty of physical phenomena. These aspects are particularly relevant in the quantum domain, where great effort has been put into modelling, studying, and designing efficient engines, whereas a general and consistent theory of entropy production would also contribute to a more quantitative understanding of irreversibility, as embedded in the open quantum systems formalism~\cite{Lindblad_book, Lendi:1988, Mauro_thermo2018, Mauro_review}. 

Given a standard open system scenario, how does the initial preparation of the system affect the entropy production rate? There are indeed different ways to tackle this multifaceted issue. For instance, one can assess the role of classical and/or non-classical correlations, either limited to those created within a composite system, or those shared between the system and the environment.
Specifically, in Ref. \cite{Zicari:2020} some numerical and analytical evidence has been gathered to prove that, whenever one considers bipartite systems, the initial entanglement shared by the two subparties play a relevant role in the entropy production rate: the maximum of the entropy production rate is algebraically dependent on the entanglement one inputs. Those results, obtained both for Markovian and non-Markovian evolutions, are derived by considering Gaussian states undergoing a Gaussian dynamics. Under these assumptions, one can always associate a well-defined probability distribution over the phase space~\cite{Ferraro:2005, Serafini:2017}. This remarkable advantage carries over to the thermodynamics of open systems --- an analysis of the entropy production in terms of Wigner or, equivalently, R\'{e}nyi-$2$ entropies is indeed suitable to the study of this scenario \cite{SantosPRL:2017}.
The analysis of harmonic systems in terms of the Wigner entropy production replaces the one based on the usual von-Neumann relative entropy. One can thus also study the case of a zero-temperature thermal bath, where one observes the so-called \emph{zero-temperature catastrophe}: in such a case, the reservoir being in a pure state, the von-Neumann relative entropy would diverge~\cite{Abe:2003,Audenaert:2014}. This is just an inconsistency of the theory, as the zero-temperature limit can be achieved in quantum optics settings, where the dynamical equations correctly reproduce experimental data~\cite{Brunelli:2018, Kewming2022entropyproduction}.
However, a phase-space description of the quantum dynamics is not limited to continuous variables systems; it can indeed be extended to spin systems, where, in order to obtain a mathematically consistent theory, the Wigner function and entropy need to be replaced by the Husimi-Q function and Wehrl entropy, respectively~\cite{Husimi:1940,Takahashi:1985,Wehrl:1978,Wehrl:1979}.

In this work, we employ a special class of states for which a space-phase description is available, namely spin coherent states~\cite{Radcliffe:1971}. Specifically, we refer to the spin-space-phase entropy production framework put forth in Ref. \cite{SantosPRA:2018}, which can be used to study the irreversibility of the Lindbladian dynamics undergone by spin systems. 

In this work, we give one more contribution to assess the role played by quantum coherence in determining relevant thermodynamic quantities~\cite{Francica:2019}, such as entropy production. For example, in Ref.~\cite{VanVu:2022}, it has been shown that -- for open systems undergoing a Lindbladian dynamics -- a mathematical bound to the entropy production can be found, giving a formal justification to the evidence that quantum coherence induces additional dissipation compared to classical protocols. 
Coherence represents an essential resource for quantum processes, setting classical and quantum phenomena apart~\cite{Coherence_review:2017}; in the context of quantum thermodynamics, the aim is essentially to ascertain whether it is either resourceful or detrimental to certain thermodynamic tasks. 
In particular, when dealing with nonequilibrium processes, coherence contributions need to be included to extend fluctuations relations in the quantum domain~\cite{Campisi:2011,Aberg:2018,Kwon:2019}. Recently, in Refs.~\cite{Hernandez_Gomez:2022,Gherardini:2021}, a protocol has been introduced to experimentally quantify the contribution to nonequilibrium entropy production stemming from the degree of quantum coherence contained in the initial state. The latter follows from the possibility to separate, under rather general conditions, the overall entropy production rate into two distinct contributions, one of which, directly related to coherence, is genuinely quantum~\cite{Landi:19}. In particular, we will focus on the case of the so-called Davies-Lindblad dynamical maps, where the splitting between classical and quantum contributions to the entropy production rate mirrors a similar separation at the level of the dynamical equations. It is indeed known that the evolution of the populations is governed by a classical Pauli master equation, whereas coherences obey a separate set of differential equations~\cite{Breuer-Petruccione}. 

Our work is meant to corroborate this statement regarding the nature of the entropy production rate with a systematic study of the interplay between the latter and quantum coherence. By considering relevant examples of dynamical maps, i.e. dephasing and amplitude damping channels, we systematically assess how different preparations of the initial state, namely different values of the initial coherence, affect the resulting spin-phase entropy production rate.  More specifically, in  similar spirit to the work reported in Ref.~\cite{Zicari:2020}, we will choose the case of a bipartite system, where the interaction between the two parts is always assumed to be null. Such a bipartite structure might also be useful to discuss the role played by non-classical correlations shared by the two parties of the system. More generally, we deliberately discard those elements that would alter the transparency of the message that we would like to deliver, such as correlations dynamically created by the process, as well as those shared by the system and the environment, or finite-bath effects~\cite{Esposito:2010,Popovic:2018,Ptaszynski:2019}. 

The remainder of this paper is organised as follows. In Sec.~\ref{sec:coherent_states} we present the apparatus used to quantify entropy production through a phase-space formalism. Sec.~\ref{sec:entropy_prod} addresses the quantification of entropy production in a compound of two two-level systems exposed to bipartite quantum channels, specifically an amplitude damping and a dephasing .

\section{Coherent states and Wehrl entropy}
\label{sec:coherent_states}

Let us consider the case of a single system, described in terms of the spin operators $J_x$, $J_y$ and $J_z$, satisfying the usual algebra $[J_x, J_y] = i J_z$ (we assume units such that $\hbar=1$ throughout the paper). A \emph{coherent spin state} is defined as~\cite{Radcliffe:1971,Arecchi:1972,Scully:1994}
\begin{align}
\label{eq:spin_coherent_def}
\ket{\Omega} = e^{- i \phi J_z} e^{- i \theta J_y} \ket{J,J},
\end{align} 
where $\ket{J,J}$ is the angular momentum state with the largest quantum number of $J_z$, while $(\phi, \theta)$ are the Euler angles. Given that a spin coherent state represents the closest quantum analog of a point in a sphere of radius $J$, we can easily explain the rationale behind the definition above. Bearing in mind the role played by Euler angles in the theory of rotations, \Cref{eq:spin_coherent_def} can be interpreted as follows: in order to reach an arbitrary point on a unit sphere, we start with the $z$ axis, then we perform a rotation around the $y$ axis by an angle $\theta$, then a second one around $z$ by $\phi$.
A straightforward calculation corroborates this analogy: one can easily show that  $
\left (\langle  J_x \rangle, \langle  J_y \rangle, \langle  J_z \rangle \right ) = \left ( J \cos{\phi} \sin{\theta}, J \sin{\phi} \sin{\theta},  J \cos{\theta} \right )
$, where $\langle \cdot \rangle \equiv \bra{\Omega} \cdot \ket{\Omega}$, so that the expectation values of the spin operators $\langle J_k\rangle~(k=x,y,z)$ embody the coordinates of a point on the surface of a sphere of radius $J$.
From \Cref{eq:spin_coherent_def}, we define the Husimi-Q function 
\begin{align}
\label{eq:Q_def}
\mathcal{Q} (\Omega) \equiv \bra{\Omega} \rho \ket{\Omega},
\end{align}
where $\rho$ is the density operator describing the system state. 
Let us assume that the system dynamics is governed by the master equation
\begin{align}
\label{eq:ME}
\dot{\rho} = - i  [ H, \rho ] +  D(\rho),
\end{align}
where $H$ is the system Hamiltonian, while $D(\rho)$ is the dissipator associated with the open dynamics, possibly in Lindblad form \cite{Breuer-Petruccione,Lindblad:1976,Gorini:1976,Gorini:1978}.
By using the Husimi-Q function, \Cref{eq:ME} in replaced by a Fokker-Planck equation for $\mathcal{Q}$, which reads as
\begin{align}
\label{eq:ME_Q}
\partial_t \mathcal{Q} = \mathcal{U} (\mathcal{Q}) + \mathcal{D} (\mathcal{Q}),
\end{align}
where $\mathcal{U} (\mathcal{Q}) $ describes the unitary part of the evolution, while $\mathcal{D} (\mathcal{Q})$ represents the dissipative part.
Given a master equation in the rather general form of \Cref{eq:ME}, one can remap it in the form of \Cref{eq:ME_Q} just using a set of correspondence rules listed in \Cref{app:phasespace}.

The Husimi-Q function features the property of being positive-definite; the latter makes the definition of the Wehrl entropy well-posed~\cite{Wehrl:1978,Wehrl:1979}:
\begin{align}
\label{eq:Wehrl_def}
S_{\mathcal{Q}} = - \left (\frac{2J +1}{4 \pi}  \right )^2 \int d \Omega \; \mathcal{Q} (\Omega) \ln{\mathcal{Q} (\Omega)},
\end{align}
where the numerical prefactor is chosen for convenience, taking into account that in this work we will deal with bipartite systems. Unlike von Neumann and Wigner entropies, the Wehrl entropy does not represent a measure of the purity of the state; it is instead directly related to the uncertainty area of the Husimi function in the phase-space~\cite{Knight:1995_1,Knight:1995_2}. This different interpretation is also related to yet another property of the Wehrl entropy: the latter, unlike the von Neumann entropy, is not invariant under unitary transformations~\cite{Wehrl:1979}.

If we take the time derivative of \Cref{eq:Wehrl_def}, together with the normalisation condition, and \Cref{eq:ME_Q}, we obtain
\begin{align}
\label{eq:rates_diss_gen}
\frac{d S_{\mathcal{Q}}}{d t} \biggl \vert_{\rm diss} =  - \left (\frac{2J +1}{4 \pi}  \right )^2 \int d \Omega \; \mathcal{D} (\mathcal{Q}) \ln{\mathcal{Q}},
\end{align}
where we consider the contribution coming from the dissipative part only. The aim is now to rewrite the latter in the Prigogine form \cite{Prigogine:1955, DeGrootMazur,Mauro_review}:
\begin{align}
\label{eq:piphi}
\frac{d S_{\mathcal{Q}}}{d t} \biggl \vert_{\rm diss} = \Pi(t) - \Phi(t),
\end{align}
where we separate the entropy flux rate $\Phi(t)$ from the entropy intrinsically produced by the process, i.e. the entropy production rate $\Pi(t)$. Note that, working in the Born-Markov approximation leading to a Lindblad master equation, one has $\Pi \ge 0$, as required by the second law of thermodynamics. On the contrary, $\Phi$ can be either positive or negative, meaning that there might be instances in which the overall entropy decreases.

\section{Entropy Production Rate}
\label{sec:entropy_prod}

In this work, we would like to study the entropy production rate in bipartite systems described in terms of spin coherent states. We thus need to consider a specific kind of dynamics for our system: we will first consider the case of dephasing channels, then we will also consider the physical situation in which termalisation occurs, i.e. amplitude damping channels~\cite{nielsen_chuang:2010}.

\subsection{Dephasing channels}
\label{sec:dephasing}

The first type of dynamics we would like to consider is represented by dephasing channels. 
We could, in principle, start from a microscopic modelling of the system-environment interaction. For example, one possibility would be to consider specific system-environment coupling in a spin-boson model so that the Hamiltonian exhibits an explicit symmetry~\cite{Breuer-Petruccione, Palma:96}. In this case, because of the lack of a direct exchange of energy between the system and the bosonic bath, one would anticipate a trivial thermodynamic behaviour. However, this is not the case, as even pure decoherence processes features rich physics, as acknowledged in Ref.~\cite{Popovic:2021}. Differently, if we restrict ourselves to the case of qubits, the same purely dephasing dynamics can emerge from a different \emph{ab-initio} derivation, e.g., the so-called shallow-pocket model~\cite{Piilo_exp:2011,Guarnieri:2014,Arenz:2015}, where the open system is represented by the two polarisation degrees of freedom of a photon, while the environment is identified with the frequency degrees of freedom. Since in such a model the environment is often assumed to be in a pure state, an analysis in terms of von-Neumann entropy would be inconsistent. 

However, regardless of the physical origin of the pure dephasing process, we will start directly from a given dissipator. In other words, for a bipartite system made of two non-interacting systems, the dissipator reads as
\begin{align}
\label{eq:deph_2spins}
D (\rho) & = D_a (\rho) +  D_b (\rho) \nonumber \\
& = -\frac{\lambda_a}{2} \left [ J_z^a, \left [ J_z^a, \rho \right ] \right ]  -\frac{\lambda_b}{2} \left [ J_z^b, \left [ J_z^b, \rho \right ] \right ],
\end{align}
where $J_z^a = J_z \otimes \mathbbm{1}_b$ and $J_z^b =  \mathbbm{1}_a \otimes J_z $, where $\mathbbm{1}_a$ and $\mathbbm{1}_b$ are the identity operators defined over the Hilbert spaces $\mathcal{H}_a$ and $\mathcal{H}_b$, associated with the first and the second spin, respectively.
Working along the same lines as in Ref.~\cite{SantosPRA:2018}, the phase-space operator correspondences given in \Cref{app:phasespace} yield the following phase-space dissipator
\begin{align}
\label{eq:deph_Q_2spins}
\mathcal{D} (\mathcal{Q}) = - \frac{\lambda_a}{2} \mathcal{J}_z^a \left ( \mathcal{J}_z^a (\mathcal{Q}) \right )  - \frac{\lambda_b}{2} \mathcal{J}_z^b \left ( \mathcal{J}_z^b (\mathcal{Q}) \right ),
\end{align}
where $\mathcal{J}_z^j (\mathcal{Q}) = - i \partial_{\phi_j} \mathcal{Q}$ $(j=a,b)$.
By plugging \Cref{eq:deph_Q_2spins} into \Cref{eq:rates_diss_gen}, after integrating by parts, and assuming that the integrand decreases sufficiently fast outside the integration volume, we can eventually identify the expression of the entropy production rate, i.e.
\begin{align}
\label{eq:entropy_prod_2quibits_deph}
\Pi \equiv \frac{d S}{d t} \biggl \vert_{\rm diss} = \Pi_a + \Pi_b, 
\end{align}
where
\begin{align}
\label{eq:entropy_prod_2quibits_deph_2}
\Pi_j \equiv \frac{ \lambda_j}{2} \left (
\frac{2J+1}{4 \pi} \right )^2 \int d \Omega \; \frac{\left | \mathcal{J}_z^j (\mathcal{Q}) \right |^2}{\mathcal{Q}}, \quad j=a,b.
\end{align}
Since the two spin systems dissipate independently, \Cref{eq:entropy_prod_2quibits_deph,eq:entropy_prod_2quibits_deph_2} yield a similar expression for the entropy production rate to the single channel case, given in \cite{SantosPRA:2018}, while correlations are encoded in the Husimi-Q function $\mathcal{Q}$ and in the integration volume  $d \Omega = \sin{\theta_a}  \sin{\theta_b} d \theta_a d \theta_b d \phi_ad \phi_b$.

\subsection{Amplitude damping channels}
\label{sec:ampl_damp}

Let us consider a more physical scenario, where the system-environment coupling leads the open system to thermalisation. Under the usual weak coupling assumption, one expects that the undriven composite system would relax towards the canonical Gibbs state given by $\rho_{\rm eq} = e^{- \beta H_{\rm S}}/ \mathcal{Z}$, where $\mathcal{Z}= \operatorname{Tr}e^{- \beta H_{\rm S}}$ is the partition function of the system, and $H_{\rm S}$ is its free Hamiltonian \cite{Breuer-Petruccione,Spohn:1977}. For instance, one can consider the case of an amplitude damping dynamics, where the dynamical process adjusts the populations to values dictated by the bath, while involving the incoherent excitations exchange between the system and the environment. The latter can be microscopically derived assuming a standard weak-coupling and memoryless scenario of a spin system interacting with a thermal bosonic reservoir, a scenario similar to the one yielding the standard quantum optical master equation for a two-level system~\cite{Breuer-Petruccione,Elouard:2020}.

In our study, we consider the case of two independent amplitude damping channels, where the dissipator of the composite system reads as
\begin{align}
D(\rho) = D_a(\rho) + D_b(\rho),
\end{align}
with
\begin{align}
\label{eq:diss_amp_dam}
D_j(\rho) = \Gamma_j  \left ( \bar{n}_j + 1\right ) & \left [ J_{-}^j \rho J_{+}^j - \frac{1}{2} \left \{  J_{+}^j J_{-}^j, \rho \right \}\right ] \nonumber \\
 + \Gamma_j \bar{n}_j & \left [ J_{+}^j \rho J_{-}^j - \frac{1}{2} \left \{  J_{-}^j J_{+}^j, \rho \right \}\right ],
\end{align}
where $j=a,b$, while $\Gamma_j$ and $\bar{n}_j$ represent the damping rate and the average number of excitations of the local reservoir, respectively. In addition, the raising/lowering spin operators for each of the two subsystems are given by $J_{\pm}^a= J_{\pm} \otimes \mathbbm{1}_b$ and $J_{\pm}^b=  \mathbbm{1}_a \otimes J_{\pm}$, where $J_{\pm} \equiv J_x \pm i  J_y$. Alternatively, the dissipators can be brought to the following form:
\begin{align}
\label{eq:diss_amp_dam_2}
D_{j}(\rho) = \frac{\Gamma_j}{2} \left \{ \left [ J_{-}^j, f_j(\rho) \right ] - \left [J_{+}^j, f_j^{\dagger} (\rho) \right ] \right \},
\end{align}
where
\begin{align}
f_j(\rho) = \left ( \bar{n}_j +1\right) \rho  J_{+}^j - \bar{n}_j J_{+}^j \rho.
\end{align}
The phase-space representation of this dissipator is given by
\begin{align}
\mathcal{D} (\mathcal{Q}) = \mathcal{D}_a (\mathcal{Q}) +  \mathcal{D}_b (\mathcal{Q}),
\end{align}
where
\begin{align}
\label{eq:diss_amp_dam_Q2}
\mathcal{D}_j (\mathcal{Q}) = \frac{\Gamma_j}{2} \left \{ \mathcal{J}_{-}^j \left (f_j(\mathcal{Q}) \right ) - \mathcal{J}_{+}^j \left (f_j^{*}(\mathcal{Q}) \right ) \right \},
\end{align}
and
\begin{align}
\label{eq:f}
f_j(\mathcal{Q})&=  \frac{1}{2} \left [ 2 J \mathcal{Q} - \mathcal{J}_z^j (\mathcal{Q}) \right ] e^{i \phi_j} \sin{\theta_j} \nonumber \\
& + \frac{1}{2} \left [ \cos{\theta_j} - \left ( 2 \bar{n}_j +1\right )\right ] \mathcal{J}_{+}^j(\mathcal{Q}).
\end{align}
Integration by parts of  \Cref{eq:rates_diss_gen}, leads to
\begin{equation}
\begin{aligned}
\frac{d S}{d t} \biggl \vert_{\rm diss} &=  \left (\frac{2J +1}{4 \pi} \right)^2 \sum_{j=a,b} \frac{\Gamma_j}{2}\int  \frac{{\cal F}_j({\cal Q})}{\mathcal{Q}} {d \Omega}
\end{aligned}
\end{equation}
with ${\cal F}_j({\cal Q})=f_j(\mathcal{Q}) \mathcal{J}_{-}^j \left ( \mathcal{Q}\right ) -  f_j^{*}(\mathcal{Q}) \mathcal{J}_{+}^j \left( \mathcal{Q} \right)$.
Since the operators of the two sub-(phase)-spaces do not mix, we can split the two contributions in a similar way as in the case of a single channel, provided that the Husimi function $\mathcal{Q}$ and the integration measure $d \Omega$ refers to the whole composite system. Therefore, we can resort to the splitting introduced in Ref.~\cite{SantosPRA:2018}, where, following standard thermodynamic arguments, the entropy production and the flux rates should be even and odd functions of the relevant currents, respectively. Extending the results presented in Ref.~\cite{SantosPRA:2018} to a bipartite spin system, on one hand we get the following expression for the entropy production rate
\begin{align}
\label{eq:Pi_amp_damp}
\Pi = \Pi_a + \Pi_b,
\end{align}
where
\begin{align}
\label{eq:Pi_amp_damp_1}
\Pi_j  &=  \left ( \frac{2J +1}{4 \pi}  \right)^2  \nonumber \\
& \times \biggl [ \frac{\Gamma_j}{2} \int \frac{d \Omega}{\mathcal{Q}}  \biggl \{  \frac{\left \{ 2 J \mathcal{Q} \sin{\theta_j} + [\cos{\theta_j} - \left ( 2 \bar{n}_j+1 \right )] \partial_{\theta_j} \mathcal{Q} \right \}^2}{\left ( 2 \bar{n}_j+1 \right ) - \cos{\theta_j}} \nonumber \\
& + \left | \mathcal{J}_z^j (\mathcal{Q}) \right |^2 \left [(2 \bar{n}_j +1) \cos{\theta_j} -1 \right ] \frac{\cos{\theta_j}}{\sin^2{\theta
_j}}\biggl \} \biggl ],
\end{align}
on the other hand, we identify the expression for the entropy flux rate, with witch each subsystem is exchanging entropy with its local thermal bath, as $\Phi = \Phi_a + \Phi_b$, where
\begin{align}
\label{eq:Phi_amp_damp_1}
\Phi_j = & \frac{J(2J{+}1)^2\Gamma_j}{16 \pi^2}  \nonumber 
\int {d \Omega}\! \left \{ \frac{2 J \mathcal{Q} \sin^2{\theta_j}}{(2 \bar{n}_j +1){-} \cos{\theta_j}}{-} \sin{\theta_j}\partial_{\theta_j} \mathcal{Q} \right \},
\end{align}
with $j=a,b$.
Similarly to the case of a single spin system~\cite{SantosPRA:2018}, the phase-space expression of the entropy production rate -- \Cref{eq:Pi_amp_damp_1} -- essentially contains two contribution: one proportional to the dephasing currents $\left | \mathcal{J}_z^j (\mathcal{Q}) \right |^2$, thus capturing the loss of coherence, the other related to the amplitude damping.

\section{Analysis and results}
Given the closed expression for the entropy production rate for both for the dephasing and amplitude damping channels, now we can discuss in depth the influence of the initial coherence on the entropy production rate. Let us suppose that at $t=0$ the composite system's density operator reads as $\rho = \xi + \chi$, where $\xi$ is the diagonal part containing populations, whereas $\chi$ containing coherences. This splitting allows us to compute the initial coherence in terms of the $l_1$-norm $\mathcal{C}(\rho) \equiv \sum_{i,j} | \chi_{ij} |$~\cite{Baumgratz:2014}. Moreover, for each channel we can regroup coherences in different classes: the elements belonging to a given class obey the same dynamical laws. 

\subsection{Two qubits: dephasing channels}
\label{sec:2qubits_deph}
Let us consider the case of two independent dephasing baths. We have $J_z^a \equiv \sigma_z^a/2$ and $J_z^b \equiv \sigma_z^b/2$, therefore the dissipator \Cref{eq:deph_2spins} reads as
\begin{equation}
\label{eq:deph_2qubits}
D (\rho)  = -\frac{\lambda}{8} \left ( \left [ \sigma_z^a , \left [ \sigma_z^a, \rho \right ] \right ]  + \left [  \sigma_z^b , \left [ \sigma_z^b , \rho \right ] \right ] \right ),
\end{equation}
where we have assumed that the dephasing rate is equal for the two channels, i.e. $\lambda_a = \lambda_b \equiv \lambda$. 

Let us denote as $\{\ket{0}_j, \ket{1}_j \}$ the basis in each two-dimensional Hilbert space $\mathcal{H}_j$, with $j=a,b$. We can thus work in the computational basis $\ket{mn} \equiv \ket{m}_a \otimes \ket{n}_b$, where $m,n \in \{0,1\}$. In order to clarify our notation, we state explicitly the form of the general density matrix that we will be considering
\begin{align}
\label{eq:2quibits_full_rho}
\rho = \begin{pmatrix}
\rho_{11} & \rho_{12} & \rho_{13} &  \rho_{14} \\
\rho^*_{12} & \rho_{22} & \rho_{23} & \rho_{24} \\
\rho^*_{13} & \rho^*_{23} & \rho_{33} & \rho_{34} \\
\rho^*_{14} & \rho^*_{24} & \rho^*_{34} & \rho_{44}
\end{pmatrix}
\end{align}
with $\rho_{44}=1-\sum^3_{i=1}\rho_{ii}$ to ensure normalization. 
We can then analytically solve the equation of motion $\dot{\rho} = D(\rho)$ in the interaction picture with respect to the free Hamiltonian of the two-qubit system. For the diagonal entries, we have
\begin{align}
\label{eq:deph_dynamics_diag}
    \rho_{ii}(t) = \rho_{ii} (0) \quad (i=1,2,3,4).
\end{align}
For the anti-diagonal entries $\rho_{14}, \rho_{23}$ and their conjugate, we have instead
\begin{align}
\label{eq:deph_dynamics_anti_diag}
    \rho_{ij}(t) = \rho_{ij}(0) e^{-\lambda t}\quad (i+j=5),
\end{align}
while the remaining entries behave as
\begin{align}
\label{eq:deph_dynamics_all_the_rest}
\rho_{ij}(t) = \rho_{ij}(0)e^{-\frac{\lambda}{2}t}.
\end{align}
\begin{figure}[t!]
\centering
\includegraphics[width=\columnwidth]{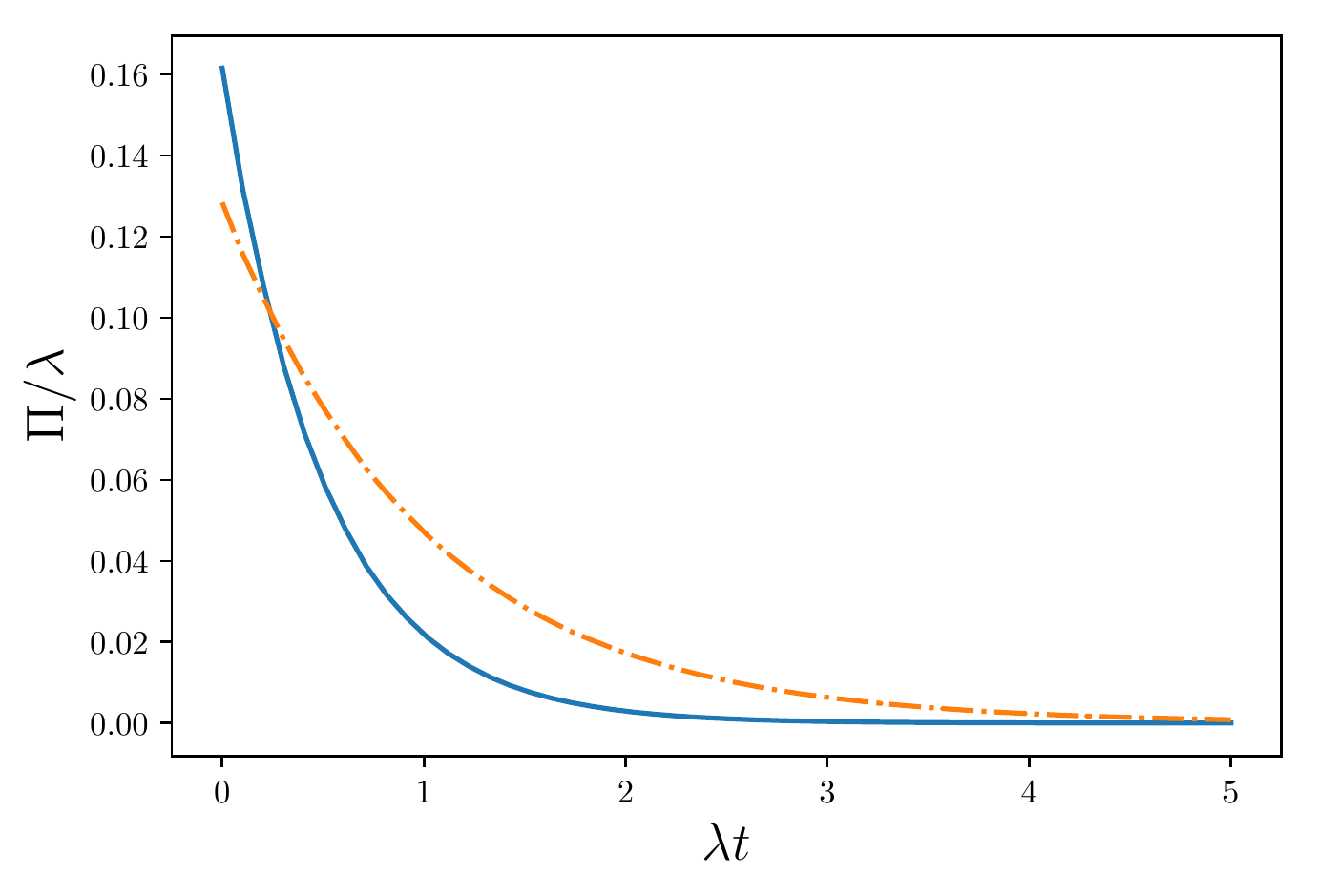}
\caption{\small{Comparison in terms of entropy production rate between X-state (solid line) and non X-states (dash-dotted line) for a given value of coherence. The former is generated by preparing the system in an initial state such that $\alpha\ne 0$ and $\beta=0$ in \Cref{eq:2quibits_deph_x_states}, while the latter is characterised by $\alpha= 0$ and $\beta\ne0$}. The system, given by two uncoupled qubits, undergoes a purely dephasing dynamics described by  \Cref{eq:deph_2qubits}. We choose $\alpha$ and $\beta$ such that the two initial states are characterised by the same value of coherence, i.e. $\mathcal{C}(\rho)=0.14$ in the numerical simulation.}
\label{fig:2qubits_deph_xstates_vs_nonx}
\end{figure}
\begin{figure*}[t!]
\centering
\includegraphics[width=2.1\columnwidth]{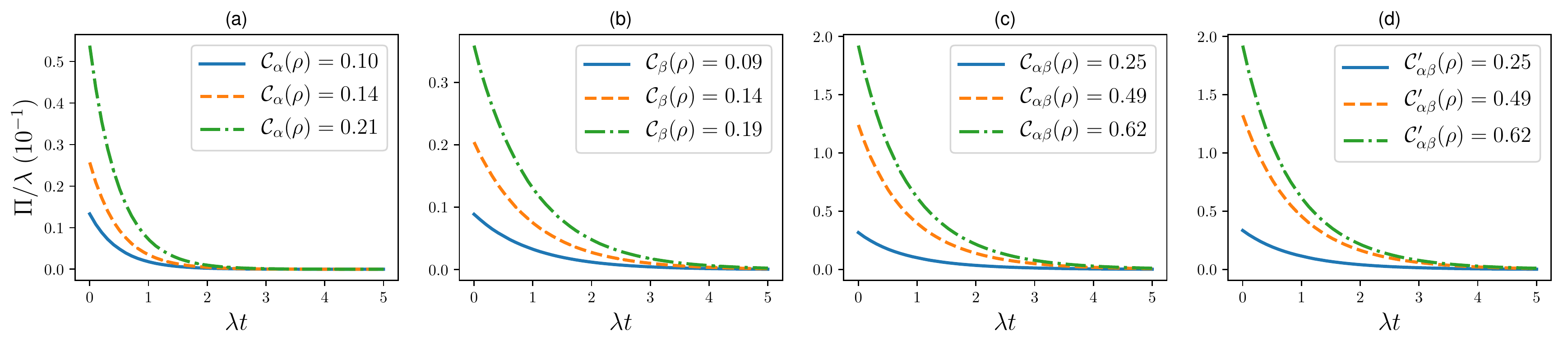}
\caption{\small{Entropy production rate as a function of time for a system of two uncoupled qubits undergoing a purely dephasing dynamics described by  \Cref{eq:deph_2qubits}. In these plots, we consider initial states in the form of \Cref{eq:2quibits_deph_x_states}. In the Panel (a), we take $\alpha\ne0$ and $\beta=0$, and we calculate the entropy production corresponding to different values of the initial coherence. The initial value of $\alpha$ is randomly taken by the uniform distribution defined over $[0,0.25]$, then we generate more initial states by rescaling coherences as described in the main text. Analogously, in the Panel (b), we choose $\beta \ne 0$ and $\alpha=0$. Finally, in the Panels (c) and (d), we choose an initial state such that $\alpha \ne 0$ and $\beta \ne 0$ simultaneously. Specifically, we choose $\alpha$ and $\beta$ from the uniform distribution on $[0,0.5]$, then, in order to generate more curves, we rescale the coherences $\alpha, \beta$ of the initial state either by the same or by different factors -- see Panels (c) and (d), respectively.}}
\label{fig:qubits_deph}
\end{figure*}
It is indeed understood that purely dephasing dynamics leaves the populations of the two qubits unchanged, while it modifies their coherences. This immediate integration of the equation of motion leads us to distinguish two different classes of coherences, each of them obeying to a given dynamical law. In this spirit, we express the density matrix of the two-qubit system in the following form
\begin{equation}
\label{eq:2quibits_deph_x_states}
\rho = \begin{pmatrix}
\rho_{11} & \beta & \beta &  \alpha \\
\beta & \rho_{22} & \alpha & \beta \\
\beta & \alpha& \rho_{33} & \beta \\
\alpha & \beta & \beta & \rho_{44}
\end{pmatrix},
\end{equation}
where we can either have the so-called X-states~\cite{Xstates:2012} when $\alpha \ne 0$ and $\beta = 0$, or non X-states when $\alpha = 0$ and $\beta \ne 0$. In the former case, the local states of each qubit are diagonal density operators that depend only on the diagonal entries of \Cref{eq:2quibits_deph_x_states}. This implies that -- in the absence of any quantum coherence -- the local states of the qubits would remain stationary throughout the dynamics. In turn, this implies that in the case of $\alpha\neq0$, entropy production stems solely from the coherences in their global state. Each class of coherence enter in a different way in the formula for the entropy production rate, whence the evidence we can gather from \Cref{fig:2qubits_deph_xstates_vs_nonx}: for a given value of coherence (in the numerical simulation we take $\mathcal{C}(\rho)= 0.14$), we obtain different curves for the entropy production rate, depending on the specific class of coherences we are assuming to be non-zero. It is worth emphasising that, given a certain initial state, the details of each curve $\Pi = \Pi(t)$ are ultimately determined by the dynamics --- Cf. \Cref{eq:deph_dynamics_anti_diag,eq:deph_dynamics_all_the_rest}. It is indeed clear that the off-diagonal terms decrease exponentially towards zero with different rates, hence curves corresponding to different classes can intersect at $t>0$, like in the case featured in \Cref{fig:2qubits_deph_xstates_vs_nonx}.

In order to study the role of initial coherence, one can prepare the composite system in one of the aforementioned classes and see what happens to the entropy production rate. In the following cases, we construct the density matrix by setting $\rho_{ii} = \xi_i/\sum_i \xi_i$, where $\xi_i$ $(i=1,2,3,4)$ is randomly chosen from the uniform distribution defined over the interval $[0,1]$. The latter condition ensures the proper normalisation of the density operator, i.e. $\operatorname{Tr}{\rho}=1$.

Let us prepare the system in an X-state, and let it evolve according to a purely dephasing dynamics given by \Cref{eq:deph_2qubits}. It is immediate to quantify the initial coherence as $\mathcal{C}_\alpha(\rho)= 4 |\alpha|$. For our numerical simulations we extracted a random value of $\alpha$ from a uniform distribution on -- for the sake of definiteness -- the interval $[0,0.25]$. Moreover, given that the density operator $\rho$ is Hermitian and normalized by construction, we need to check that it is positive semi-definite, i.e. its eigenvalues are all required to be non-negative.  

We can then keep the populations constant, while we rescale the coherences in such a way that $\alpha \to \mu \alpha$, $\mu$ being a suitable scaling constant; under this hypothesis, it is not difficult to show that the entropy production rate would scale as $\Pi \to |\mu|^2 \Pi$.
Therefore, one would anticipate that, by increasing the coherence, one should get a higher entropy production rate. We got evidence that this in the plots shown in the Panel (a) of \Cref{fig:qubits_deph}, where one can notice that a higher coherence in terms of $l_1$-norm results in a higher entropy production rate. For each time-step, the value of the entropy production rate is computed by performing a Monte-Carlo integration of \Cref{eq:entropy_prod_2quibits_deph_2}. Similarly, one can initialise the system in a non-X state, where the initial coherence is given by  $\mathcal{C}_\beta(\rho)= 8 |\beta|$. By extracting $\beta$ from the uniform distribution defined over $[0,0.25]$, one can rescale the coherences as in the previous case, so that we get an analogous scaling law for the entropy production rate [Cf. \Cref{fig:qubits_deph}(b)].

Finally, we can combine the two classes of initial states by taking $\alpha \ne 0$ and $\beta \ne 0$ in \Cref{eq:2quibits_deph_x_states}, with the initial coherence being given by $\mathcal{C}_{\alpha \beta}(\rho)= 4 (|\alpha|+ 2 |\beta|)$. In other terms, we are mixing coherences whose time-evolutions are governed by different dynamical laws --- see \Cref{eq:deph_dynamics_anti_diag,eq:deph_dynamics_all_the_rest}. In this case, if we rescale the two classes of coherences by the same factor $\mu$, we would still recover the same scaling law as in the previous cases, i.e. $\Pi \to |\mu|^2 \Pi$. Numerically, this is shown in \Cref{fig:qubits_deph}(c). In general, if we rescale them differently, i.e. $\alpha \to \mu \alpha$  and $\beta \to \mu' \beta$, we would not obtain the same analytical result for the entropy production rate. Nevertheless, we can still numerically assess that preparing the initial state with a higher coherence will give a higher entropy production rate, as shown in \Cref{fig:qubits_deph}(d). Note that, in the numerical simulations of Panels (c) and (d), the values of $\alpha$ and $\beta$ from the uniform distribution on the interval $[0,0.5]$.

\subsection{Two qubits: amplitude-damping channels}
\label{sec:2qubits_ampl_damp}

We can now consider the case of two independent qubits undergoing an amplitude damping dynamics. The dissipator in \Cref{eq:diss_amp_dam_2} becomes
\begin{align}
\label{eq:ampl_damp_2qubits}
D (\rho)  = D_a (\rho) + D_b (\rho),
\end{align}
where
\begin{equation}
\begin{aligned}
    D_j(\rho) &= \Gamma \left ( \bar{n} + 1\right )  \left [ \sigma_{-}^j \rho \sigma_{+}^j - \frac{1}{2} \left \{  \sigma_{+}^j \sigma_{-}^j, \rho \right \}\right ] \\
 &+ \Gamma \bar{n}  \left [ \sigma_{+}^j \rho \sigma_{-}^j - \frac{1}{2} \left \{  \sigma_{-}^j \sigma_{+}^j, \rho \right \}\right ],
\end{aligned}
\end{equation}
with $j=a,b$. We have also assumed that the two channels are characterised by the same damping rate $\Gamma$ and average number of environmental excitations $\bar{n}$. 

Working in the computational basis, we can represent the generic density operator as in \Cref{eq:2quibits_full_rho}. Given the dissipator in \Cref{eq:ampl_damp_2qubits}, one can analytically solve the set of coupled differential equations coming from $\dot{\rho} = D(\rho)$ [Cf. \Cref{app:ampl_damp_dynamics} for more details]. The explicit solution enables us to distinguish three different classes of coherences -- which we label as $\alpha, \beta, \gamma$ -- appearing in the following density matrix
\begin{align}
\label{eq:2quibits_ampl_damp_classes}
\rho = \begin{pmatrix}
\rho_{11} & \alpha & \alpha &  \gamma \\
\alpha & \rho_{22} & \gamma & \beta \\
\alpha & \gamma & \rho_{33} & \beta \\
\gamma & \beta & \beta & \rho_{44}
\end{pmatrix}.
\end{align}
\subsubsection{Von-Neumann vs Wehrl entropy}

First, we would like to compare values for the entropy production using the traditional approach based on von-Neumann relative entropy with the phase-space approach based on Wehrl entropy. To this end, we can choose the same initial state, let it evolve through the channel given by \Cref{eq:ampl_damp_2qubits}, then compute the entropy production rate either using \Cref{eq:Pi_amp_damp,eq:Pi_amp_damp_1}, or the equivalent expression in terms of von-Neumann entropy. To this end, it it useful to remind that, in the standard scenario, the entropy production rate is computed using the von-Neumann relative entropy, defined as
\begin{align}
    \label{eq:von-neumann}
    S_{\rm vN}(\rho || \rho_{\rm eq}) \equiv \operatorname{Tr} \left ( \rho \ln{\rho} - \rho \ln{\rho_{\rm eq}}\right),
\end{align}
\begin{figure}[b!]
\centering
\includegraphics[width=0.8\columnwidth]{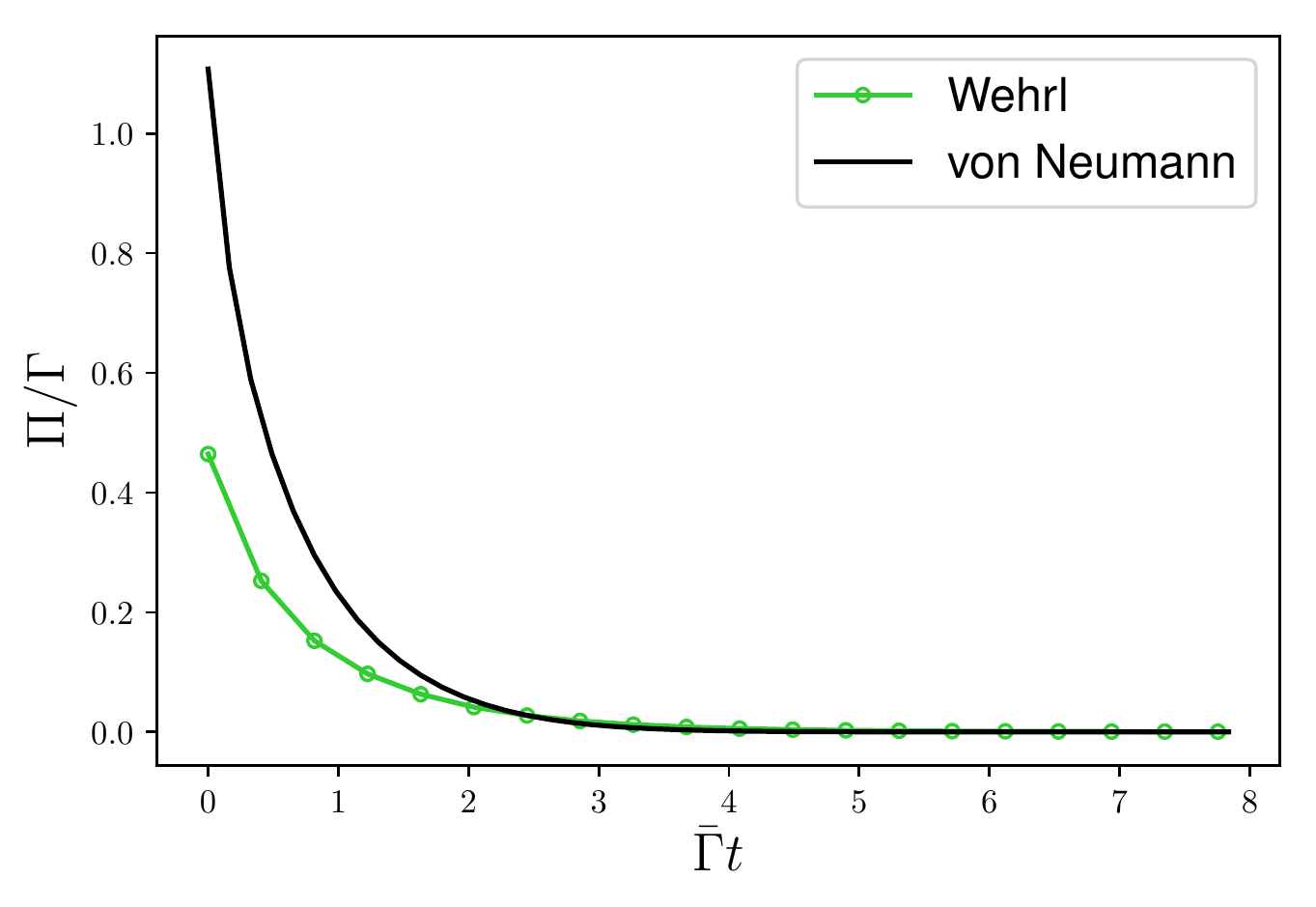}
\caption{\small{Comparison between Wehrl and von Neumann entropy production rates, i.e. \Cref{eq:Pi_amp_damp_1} or \Cref{eq:von-neumann_entropy_prod}, respectively. We prepare the bipartite system in the state given by \Cref{eq:2quibits_ampl_damp_classes}, and evolve it through \Cref{eq:ampl_damp_2qubits}. We have taken $\rho_{11,33}=0.1, \; \rho_{22}=0.2$, whereas $\alpha=\gamma=0.02$ and $\beta=0.15$. The physical parameters of the dynamics are $\bar{n}=1.5$,  $\epsilon=1$. We have set $\bar{\Gamma}= \Gamma (2 \bar{n} + 1)$.}}
\label{fig:Wehrl_vs_vN}
\end{figure}
\begin{figure*}[t!]
\centering
\includegraphics[width=2.1\columnwidth]{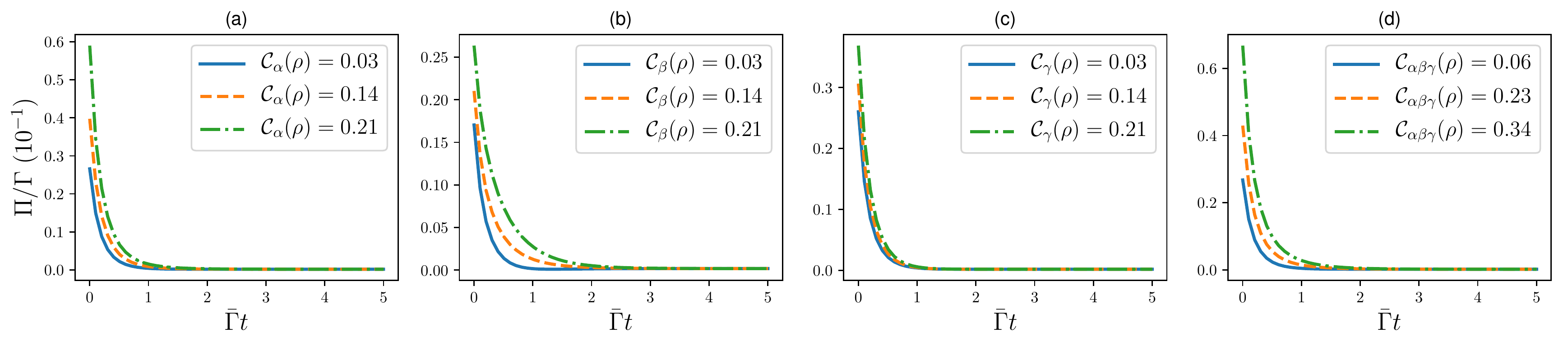}
\caption{\small{Entropy production rate as a function of time for a system of two uncoupled qubits whose dynamics follows the amplitude damping channel given by \Cref{eq:ampl_damp_2qubits}. In these plots, we consider initial states in the form of \Cref{eq:2quibits_ampl_damp_classes}. In the Panel (a), we take $\alpha\ne0$ and $\beta, \gamma =0$, and we calculate the entropy production corresponding to different values of the initial coherence. Once we construct a physical state by randomly choosing $\alpha$ from the uniform distribution on the interval $[0,0.25]$, we can generate more states just by rescaling the initial coherence. Analogously, in the Panel (b) and (c), we consider the classes corresponding to $\beta \ne 0$ while $\alpha, \gamma =0$, and $\gamma \ne 0$ while $\alpha, \beta$, respectively. In Panel (d), the initial state is given by \Cref{eq:2quibits_ampl_damp_classes}, where all the entries are non-zero. Note that in all the numerical simulations $\alpha, \beta, \gamma$ are chosen from the uniform distribution on $[0,0.25]$, whereas the dynamics of the two-qubit system is simulated taking $\bar{n}=0.5$. As in the previous plots, we have introduced the bath-temperature dependent rate $\bar{\Gamma}= \Gamma (2 \bar{n} + 1)$.}}
\label{fig:ampl_damp}
\end{figure*}
where $\rho_{\rm eq}= e^{-\beta_{\rm R} H_{\rm S}}/ \mathcal{Z}$ is the canonical equilibrium Gibbs state ($\beta_{\rm R}$ is the thermal reservoir bath temperature), the two-qubit free Hamiltonian reads as $H_{\rm S}= \frac{\epsilon_a}{2}( \sigma_z \otimes \mathbbm{1}_2) + \frac{\epsilon_b}{2} (\mathbbm{1}_2 \otimes \sigma_z)$, and $\epsilon_j$ accounts for the splitting between the two levels of each qubit $(j=a,b)$. Within this framework~\cite{Spohn:1978}, the entropy production rate is given by
\begin{align}
    \label{eq:von-neumann_entropy_prod}
    \Pi_{\rm vN}(t) = - \frac{d}{dt} S_{\rm vN}(\rho || \rho_{\rm eq}).
\end{align}
In \Cref{fig:Wehrl_vs_vN} we compare the two curves, one corresponding to the von Neumann entropy production given by \Cref{eq:von-neumann_entropy_prod}, the other corresponding to its phase-space counterpart --- Cf. \Cref{eq:Pi_amp_damp_1}. The two expressions qualitatively reproduce a similar behaviour: the entropy production rate starts from a non-null value, then it monotonically decreases to zero until the steady state is reached. 
However, the Wehrl entropy production rate given in \Cref{eq:Pi_amp_damp_1} is process-specific, in the sense that its final expression bears dependence on the specific form of the dissipator: as discussed in \Cref{sec:ampl_damp}, one can distinguish different phase-space currents corresponding to contributions coming from different physical processes underlying the overall dynamics, i.e. dephasing and amplitude damping. However, the evaluation of the spin-phase-space entropy production poses the additional technical difficulty embodied by the need to integrate over the whole phase space. Details about the parameters used for the simulations are given in the caption of \Cref{fig:Wehrl_vs_vN}.

\subsubsection{Influence of the initial coherence}

In analogy with the case of dephasing channels discussed in \Cref{sec:2qubits_deph}, the idea is to systematically prepare the system in a state belonging to one of the aforementioned classes, checking how different preparations of the initial state affect to the entropy production rate. Since the expression for the entropy production rate is much more complicated than the one corresponding to the pure-dephasing dynamics -- as one can immediately deduce by comparing \Cref{eq:Pi_amp_damp_1,eq:entropy_prod_2quibits_deph_2} -- it is not possible to obtain an analytical scaling law in such a case, but we need to proceed by numerical investigations only. 

Let us consider the case in which the initial state is given by \Cref{eq:2quibits_ampl_damp_classes}, with $\alpha \ne 0$, and $\beta, \gamma = 0$. Using the $l_1$-norm, we can quantify the coherence, i.e. $\mathcal{C}_\alpha(\rho)= 4 |\alpha|$. Analogously, one can choose the initial state such that $\beta \ne 0$ and $\alpha, \gamma =0$; therefore, $\mathcal{C}_\beta(\rho)= 4 |\beta|$. If we take $\gamma \ne 0$, while keeping both $\alpha$ and $\beta$ null, we obtain a X-shaped state, with the initial coherence being given by $\mathcal{C}_\gamma(\rho)= 4 |\gamma|$.
Working along the same lines as in \Cref{sec:2qubits_deph}, for each class, once generated a random initial state, we obtain more initial states of the same class just by rescaling all the coherences by the same quantity $\mu$, e.g., $\alpha \to \mu \alpha$. In all these cases, as shown in the Panels (a-c) of  \Cref{fig:ampl_damp}, we can clearly see that the higher the initial coherence is, the higher the corresponding entropy production rate. 
In principle one can push further this systematic analysis by considering all the possible combinations of non-null entries in the preparation of the initial state. Since all those cases bear similarity between them, in the Panel (d) of \Cref{fig:ampl_damp}, we only consider the case in which all the entries of the density matrix in \Cref{eq:2quibits_ampl_damp_classes} are different from zero, i.e. such that the initial coherence is given by $\mathcal{C}_{\alpha \beta \gamma}(\rho)= 4 (|\alpha| + |\beta|+|\gamma|)$. This general instance confirms that a higher amount of initial coherence yields a higher entropy production rate. Note that in all the numerical simulations, we extracted the entries $\alpha, \beta, \gamma$ from the uniform distribution defined over the interval $[0,0.25]$.

\begin{figure*}[t!]
         \includegraphics[width=\columnwidth]{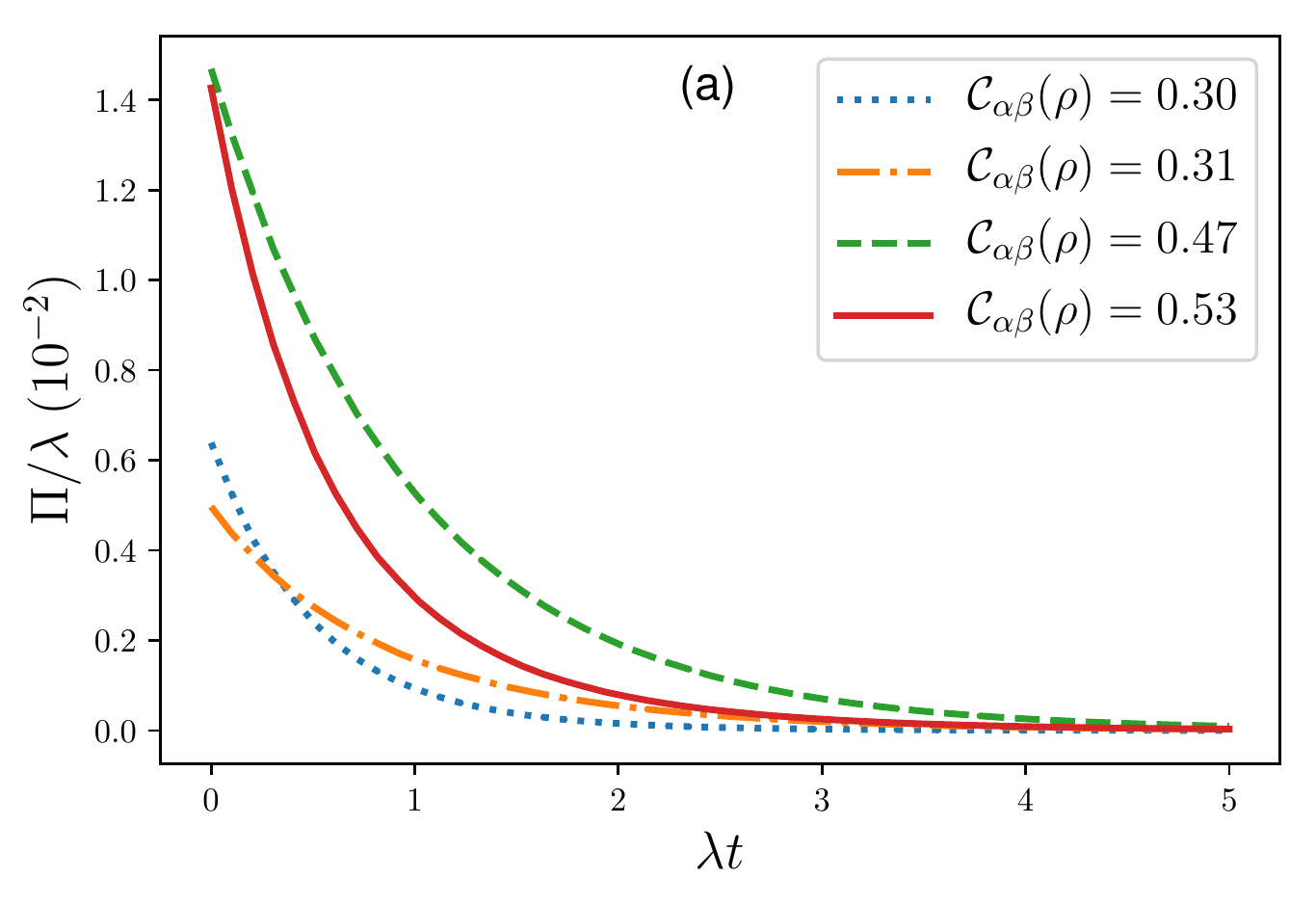}
         \includegraphics[width=\columnwidth]{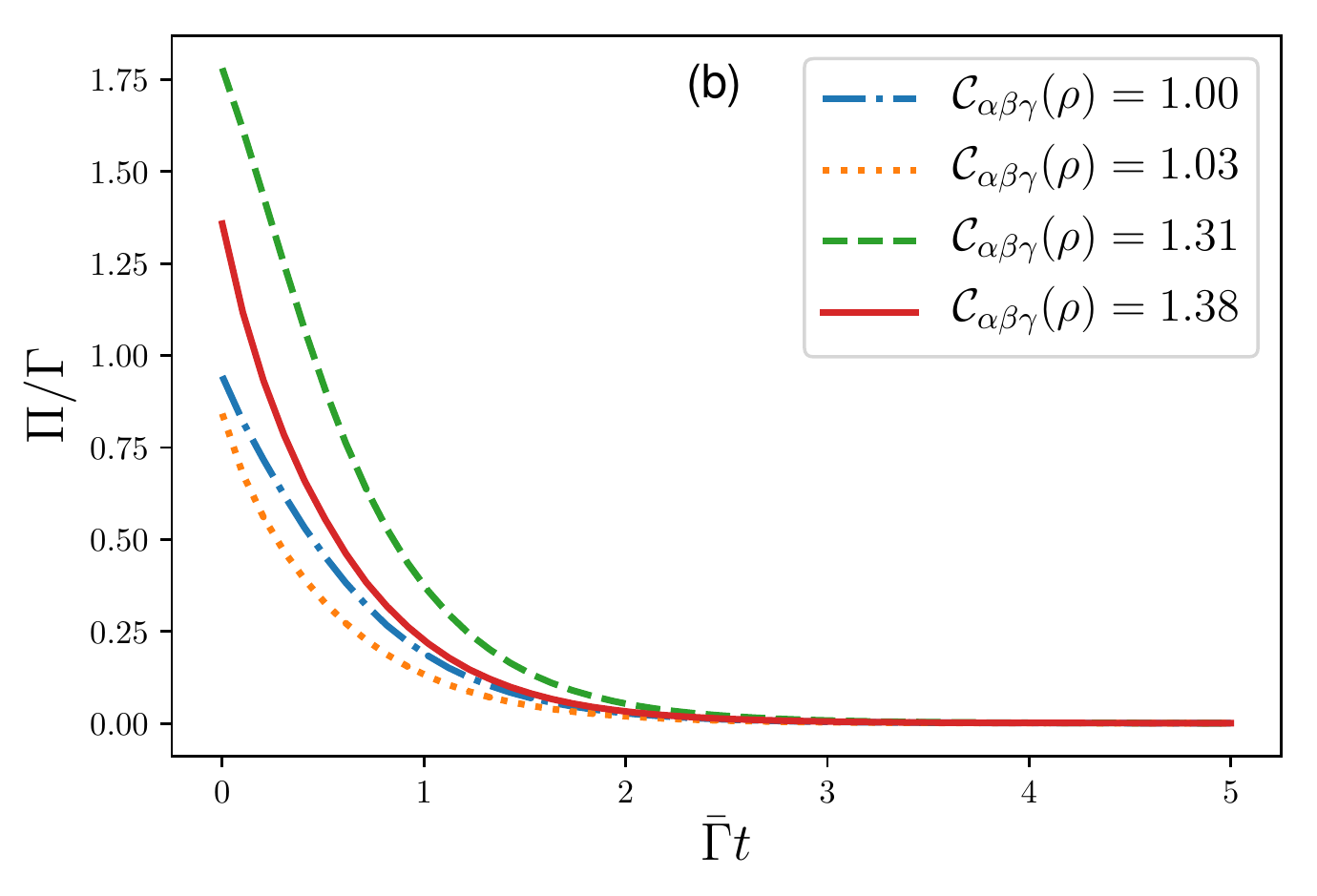}
        \caption{{These plots show that a higher values of the initial coherence does not always correspond to a higher entropy production rate. The two-qubit system is prepared in a state either in the form of \Cref{eq:2quibits_deph_x_states} (Panel (a)) or that of \Cref{eq:2quibits_ampl_damp_classes} (Panel (b)), then it undergoes a dissipative dynamics governed by \Cref{eq:deph_2qubits} or \Cref{eq:ampl_damp_2qubits}, respectively. Different initial states are considered by randomly choosing $\alpha, \beta, \gamma$ from the uniform distribution on the interval $[0,0.25]$.  Moreover, in Panel (b), $\bar{\Gamma}= \Gamma (2 \bar{n} + 1)$, and the dynamics of the two-qubit system is simulated taking $\bar{n}=0.5$.}}
        \label{fig:counterexamples}
\end{figure*}
\subsection{Further remarks and counterexamples}

The analysis presented in \Cref{sec:2qubits_deph,sec:2qubits_ampl_damp} relies on quantifying the degree of coherence contained in the initial state through the $l_1$-norm. However, our findings continue to hold also when we rely on entropic quantifiers, such as the relative entropy of coherence~\cite{Baumgratz:2014}, given by
\begin{align}
\label{eq:relative_coh}
\mathcal{C}_{\rm rel}(\rho) = S(\xi) - S(\rho),
\end{align}
where we refer to the decomposition $\rho = \xi + \chi$ of the initial density matrix, while $S(\rho) \equiv - \operatorname{Tr} \rho \ln \rho$ stands for the von Neumann entropy. For the class of states introduced in \Cref{sec:2qubits_deph,sec:2qubits_ampl_damp}, a monotonicity relation between the two quantifiers $\mathcal{C}(\rho)$  and $\mathcal{C}_{\rm rel}(\rho)$ holds, therefore, although the two quantifiers take different numerical values, the numerical hierarchy established in \Cref{fig:qubits_deph,fig:ampl_damp} is qualitatively preserved regardless of the specific quantifier used.

Moreover, from the examples above, one might argue that initial states characterised by a higher initial coherence are those yielding a higher entropy production rate. However, we get this evidence using a relatively restrictive class of initial states. We have generated the hierarchy of states by following a procedure where a state is randomly generated, then more states are generated by suitably rescaling coherences according to a given scaling parameter. Differently, one can explore more systematically the space of physical states by independently generating different random initial states, which can be labelled in terms of their coherence, quantified by the $l_1$-norm. In this more general scenario, when investigating the behaviour of $\Pi= \Pi(t)$, we cannot claim anymore that a higher value of the initial coherence would result in a higher entropy production rate.

To corroborate this statement, we provide numerical evidence by considering some instances in \Cref{fig:counterexamples}. The latter overall rule out the possibility to claim a general monotonicity relationship between degree of intial coherence and entropy production rate for initial states more general than those scrutinised in \Cref{sec:2qubits_deph,sec:2qubits_ampl_damp}.
In Panel (a) we consider the case of dephasing channels given by \Cref{eq:deph_2qubits}, whereas in Panel (b) we consider amplitude damping channels of \Cref{eq:ampl_damp_2qubits}. In both cases, we randomly generate the initial states, namely using \Cref{eq:2quibits_deph_x_states} or \Cref{eq:2quibits_ampl_damp_classes}, respectively; we then track the time evolution of $\Pi(t)$. The initial coherences are extracted from the uniform distribution on $[0,0.25]$. Besides, for the amplitude damping channels, we extract the populations $\rho_{ii}$ of the density matrix from the uniform distribution on $[0,1]$, as described in \Cref{sec:2qubits_deph}. This suggests that the grouping the coherences according to the dynamical equations that they obey throughout the dynamics introduces a monotonicity relationship between the $l_1$-norm and entropy production only when we fix the initial state and rescale the coherences.

\section{Conclusions}

We  have considered a finite-size bipartite quantum system out of equilibrium through local non-unitary channels, and --- using the tool embodied by the Wehrl entropy, which remains to be well-defined in the zero-temperature limit --- showed that no general hierarchy exists between  entropy production and the degree of quantum coherence in the state of the system. A direct correspondence between such quantities can only be retrieved when considering specific forms of open-system dynamics applied to suitably chosen initial states, thus providing evidence of the need for a systematic study of the role of genuine quantum features in the non-equilibrium thermodynamics of quantum processes.

\acknowledgments
We acknowledge discussions with Matteo Brunelli in the early development of the problem. GZ and B\c{C} are grateful to Steve Campbell for hospitality within his group at University College Dublin during the final stages of this project. 
We acknowledge support from the European Union's Horizon 2020 FET-Open project TEQ (Grant Agreement No. 766900), the Horizon Europe EIC Pathfinder project QuCoM (Grant Agreement No. 101046973), the Leverhulme Trust Research Project Grant UltraQuTe (grant RGP-2018-266), the Royal Society Wolfson Fellowship (RSWF/R3/183013), the UK EPSRC (grant EP/T028424/1) and the Department for the Economy Northern Ireland under the US-Ireland R\&D Partnership Programme. B\c{C} is supported by The Scientific and Technological Research Council of Turkey (TUBITAK) under Grant No.~121F246 and BAGEP Award of the Science Academy.

\appendix

\section{Phase-space correspondence rules}
\label{app:phasespace}

In analogy with the case of bosonic coherent states, one can establish a set of correspondence rules, which enable to write down a Fokker-Planck equation for the Husimi-Q function $\mathcal{Q}$ from the corresponding master equation for the density operator $\rho$. In particular, we have~\cite{SantosPRA:2018}
\begin{align}
\left[J_{+} , \rho \right ] &\rightarrow \mathcal{J}_{+} (\mathcal{Q}) = e^{i \phi} \left (\partial_{\theta} + i \cot{\theta} \; \partial_{\phi} \right ) \mathcal{Q}\\
\left[J_{-} , \rho \right ] & \rightarrow \mathcal{J}_{-} (\mathcal{Q}) = -e^{-i \phi} \left (\partial_{\theta} - i \cot{\theta} \; \partial_{\phi} \right )\mathcal{Q},\\
\left[J_{z} , \rho \right ] & \rightarrow \mathcal{J}_{z} (\mathcal{Q}) = - i {\partial}_{\phi} \mathcal{Q}.
\end{align}
It is worth mentioning that the analogy between bosonic and spin coherent states can be pushed further thanks to Schwinger's theory of angular momentum, according to which spin operators are represented by a pair of bosonic operators. As a result, once performed the so-called Takahashi-Shibata-Schwinger (TSS) mapping, one can use the formalism of standard bosonic coherent states \cite{Shibata:1975}. However, for the specific issue we would like to address, the TSS  approach would be quite cumbersome, as we would need to introduce four bosonic operators, two for each sub-party of the bipartite system.

\section{Amplitude Damping Channel}
\label{app:ampl_damp_dynamics}

In this Appendix, we give the explicit solution of the dynamics for a two-qubit system undergoing an amplitude damping dynamics. In other words, assuming that the system's dynamics is governed by the dissipator given in \Cref{eq:ampl_damp_2qubits}, one can explicitly solve the dynamics $\dot{\rho} = D(\rho)$. As one would expect for Davies-Lindlblad maps~\cite{Breuer-Petruccione}, the evolution of populations is decoupled from the one of coherences. 
Since the grouping into three different classes put forward in \Cref{sec:2qubits_ampl_damp} is based on the time evolution of coherences, we just report here the dynamical laws involving coherences. The analytical solution shows that
\begin{align}
    \rho_{12}(t) &= \frac{e^{-\frac{3}{2}\bar{\Gamma} t}}{1+ 2 \bar{n}} \big \{ (1 + \bar{n})\rho_{12}(0) - \bar{n} \rho_{34}(0)  \nonumber \\
    & + \bar{n} e^{\bar{\Gamma} t} \left ( \rho_{12}(0) + \rho_{34}(0)\right ) \big \} \\
    \rho_{34}(t) & = \frac{e^{-\frac{3}{2}\bar{\Gamma} t}}{1+ 2 \bar{n}} \big \{ \bar{n} \rho_{34}(0) - (1 + \bar{n})\rho_{12}(0) \nonumber \\
     & + (\bar{n} +1 ) e^{\bar{\Gamma} t}  \left ( \rho_{12}(0) + \rho_{34}(0)\right ) \big \}
\end{align}
where $\bar{\Gamma} = \Gamma (2 \bar{n} +1)$. A similar set of equations govern the evolution of $\rho_{13}$ and $\rho_{24}$, provided that $\rho_{12} \to \rho_{13}$ and $\rho_{34} \to \rho_{24}$. The remaining coherences exponentially decrease over time as $\rho_{ij}(t) = e^{-\bar{\Gamma}t} \rho_{ij}(0)$. These dynamical laws justify the classes introduced in \Cref{eq:2quibits_ampl_damp_classes}.

\bibliographystyle{apsrev4-1.bst}
\bibliography{biblio.bib}

\end{document}